\documentstyle[aps,12pt,preprint,psfig]{revtex}

\def \to {\rightarrow}
\def \beq {\begin{equation}}
\def \eeq {\end{equation}}
\def \ba {\begin{eqnarray}}
\def \ea {\end{eqnarray}}
\def \jpsi {J/\psi}

\def \< {\left <}
\def \> {\right >}
\def \mp {\mu^+\mu^-}

\begin{document}
\tightenlines
\preprint{
\hbox{PKU-TP-98-52}}
\draft

\title{Quark initiated coherent diffractive production of muon pair
        and $W$ boson at hadron colliders}

\author{Feng Yuan}
\address{\small {\it Department of Physics, Peking University, Beijing 100871, People's Republic
of China}}
\author{Kuang-Ta Chao}
\address{\small {\it China Center of Advanced Science and Technology (World Laboratory), Beijing 100080,
        People's Republic of China\\
      and Department of Physics, Peking University, Beijing 100871, People's Republic of China}}

\maketitle
\begin{abstract}

The large transverse momentum muon pair and $W$ boson productions in the
quark initiated coherent diffractive processes at hadron colliders
are discussed under the framework of the two-gluon exchange parametrization
of the Pomeron model. In this approach, the production cross sections are
related to the small-$x$ off-diagonal gluon distribution and the large-$x$
quark distribution in the proton (antiproton).
By approximating the off-diagonal gluon distribution by the usual gluon
distribution function, we estimate the production rates of these processes
at the Fermilab Tevatron.
\end{abstract}


\section{Introduction}
In recent years, there has been a renaissance of interest in
diffractive scattering.
These diffractive processes are described by the Regge theory in
terms of the Pomeron ($I\!\!P$) exchange\cite{pomeron}.
The Pomeron carries quantum numbers of the vacuum, so it is a colorless entity
in QCD language, which may lead to the ``rapidity gap" events in experiments.
However, the nature of the Pomeron and its interaction with hadrons remain a mystery.
For a long time it had been understood that the dynamics of the
``soft Pomeron'' was deeply tied to confinement.
However, it has been realized now that much can be learned about
QCD from the wide variety of small-$x$ and hard diffractive processes,
which are now under study theoretically and experimentally.

On the other hand, as we know that there exist nonfactorization effects
in the hard diffractive processes at hadron colliders
\cite{preqcd,collins,soper,tev}.
First, there is the so-called spectator effect\cite{soper}, which can
change the probability of the diffractive hadron emerging from collisions
intact. Practically, a suppression factor (or survive factor) "$S_F$"
is used to describe this effect\cite{survive}.
Obviously, this suppression factor can not be calculated in perturbative
QCD, and is now viewed as a nonperturbative parameter.
Typically, the suppression factor $S_F$ is determined to be about
$0.1$ at the energy scale of the Fermilab Tevatron\cite{tev}.
Another nonfactorization effect discussed in literature is the coherent
diffractive processes at hadron colliders\cite{collins}, in which
the whole Pomeron is induced in hard scattering.
It is proved in \cite{collins} that the existence of the leading twist
coherent diffractive process is associated with a breakdown of the
QCD factorization theorem.

Recently, we have shown that these coherent
diffractive processes can be studied in the two-gluon exchange model\cite{th1,th2,th3},
and have calculated the diffractive $\jpsi$ and charm jet production rates
at hadron colliders\cite{psi,charm}.
The two-gluon exchange parametrization of the Pomeron were previously
used to describe the diffractive processes in photoproduction at $ep$
colliders\cite{th1,th2,th3}.
An important feature of this model recently demonstrated is the sensitivity
of the production cross section to the off-diagonal parton distribution
functions in the proton\cite{offd}.

As sketched in Fig.1, the color-singlet two-gluon system (representing
the Pomeron) is emitted from one hadron and interacts with another hadron
by scattering off its partons in a hard scattering process. This process
is a coherent diffractive process.
The processes calculated in \cite{psi,charm} are the gluon induced processes,
i.e., the parton (emitted from the upper hadron)
involved in the hard process is the gluon.
Here, in this paper, the processes that will be calculated
are the quark initiated processes.
Following the method introduced in \cite{psi,charm}, we calculate the massive
muon pair and $W$ boson production in the coherent diffractive processes
at hadron colliders in the leading logarithmic approximation (LLA) of QCD.
As shown in Fig.1, the massive muon pair is produced via a time-like
virtual photon $\gamma^*$.
So, we only need to calculate the process of a virtual photon production,
$p\bar p \to q\gamma^* p+X$, and the virtuality of the photon $M^2$ is
equal to the invariant mass of produced muon pair.

Because of the quark participation in these processes, we expect that the
production cross sections of these processes are sensitive to the large-$x$
quark distribution in the proton. As has been shown in \cite{psi,charm},
the gluon initiated coherent processes are sensitive to the small-$x$
gluon distribution in the proton.
However, all of these coherent processes are sensitive to the off-diagonal
gluon distribution in the proton, which is the typical property of the
two-gluon exchange parametrization of the Pomeron model calculated previously
and here.

The diffractive production of heavy quark jet at hadron colliders has also
been studied by using the two-gluon exchange model in Ref.\cite{levin}.
However, their calculation method is very different from ours
\footnote{For detailed discussions and comments, please see \cite{charm}}.
In their calculations, they separated their diagrams into two parts,
and called one part the coherent diffractive contribution to the heavy
quark production.
However, this separation can not guarantee the
gauge invariance\cite{charm}.
In our approach, we follow the definition of
Ref.\cite{collins}, i.e., we call
the process in which the whole Pomeron participants in the hard scattering
process as the coherent diffractive process.
Under this definition, all of the diagrams plotted in Fig.2
for the partonic processes $qp\rightarrow q\gamma^* p$
and $qp\rightarrow q'W p$ contribute to
the coherent diffractive production.

In addition, we must emphasize that in this paper we only consider the
diffractive muon pair (or $W$) production in the ``Pomeron Fragmentation
region", i.e., $M_X^2\sim 4k_T^2$, where $M_X^2$ is the invariance mass of
the diffractive final states and $k_T$ is the transverse momentum of the 
muon pair.

The rest of the paper is organized as follows.
In Sec.II, we give the cross section formulas for the partonic processes
$qp\to \gamma^*q p$ and $qp\to W^\pm q p$ in LLA QCD.
We use the Feynman rule method\cite{wust} in the calculations.
The numerical results are given in Sec.III for diffractive massive muon pair
and $W$ boson productions at large transverse momentum
at the Fermilab Tevatron.
The conclusion is given in Sec.IV.

\section{ LLA formula for the partonic process}

The Feynman diagrams of the partonic processes are the same for the
$\gamma^*$ production and the $W$ boson production except the difference
of the electroweak vertex in these two processes.
So, in the following, we will mainly focus on the calculation of the
$\gamma^*$ production process. The $W$ boson production cross section will
be obtained by the similar method.

As sketched in Fig.1, the cross section for the diffractive muon pair production
at hadron colliders ($p\bar p$ at the Tevatron) can be formulated as
\beq
d\sigma(p\bar p\to \mp p+X)=\int dx_1d\hat{\sigma}(qp\to \mp qp)f_q(x_1,Q^2),
\eeq
where $x_1$ is the longitudinal momentum fraction of the antiproton
carried by the incident quark of flavor $q$.
$f_q(x_1,Q^2)$ is the quark distribution function in the antiproton, and 
$Q^2$ is the scale of the hard process.
$d\hat{\sigma}(qp\to \mp q p)$ is the cross section for the partonic process
$gp\to \mp q p$.

In the diffractive partonic process $qp\to
\gamma^* q p$, at the leading order of perturbative QCD, there
are four diagrams shown in Fig.2.
The two-gluon system coupled to the diffractive proton is in
a color-singlet state, which characterizes the diffractive processes in
perturbative QCD.
Due to the positive signature of these diagrams (color-singlet exchange),
the real part of the amplitude cancels out in the leading
logarithmic approximation.
To evaluate the imaginary part of the amplitude, we must calculate the
discontinuity represented by the crosses in each diagram of Fig.2.

The diagrams of Fig.2 show that the process $qp\to \gamma^* qp$ is by crossing
related to the photoproduction process $\gamma^* p\to q\bar q p$ in $ep$
collisions. But the virtualities of the photons in these two processes are not
the same. In the photoproduction process (DIS) the virtual photon is space-like,
while in the massive muon production process it is time-like.
This relation between the above two processes is similar to the relation between
the Drell-Yan process at hadron colliders and the deep inelastic scattering
process at $ep$ colliders.

In the following calculations, we express the formulas in terms of
the Sudakov variables.
That is, every four-momenta $k_i$ are decomposed as
\beq
k_i=\alpha_i q+\beta_i p+{k}_{iT},
\eeq
where $q$ and $p$ are the momenta of the incident quark and the proton,
$q^2=0$, $p^2=0$, and $2p\cdot q=W^2=s$.
Here $s$ is the c.m. energy of the quark-proton system, i.e., the invariant
mass of the partonic process $qp\to \gamma^*q p$.
$\alpha_i$ and $\beta_i$ are the momentum fractions of $q$ and $p$
respectively.
$k_{iT}$ is the transverse momentum, which satisfies
\beq
k_{iT}\cdot q=0,~~~
k_{iT}\cdot p=0.
\eeq

All of the Sudakov variables for every momentum
are determined by using the on-shell conditions
of the momenta of the external-line and crossed-line particles in the diagram.

The Sudakov variables associated with the momentum $k$ are determined by the
on-shell conditions of the outgoing $\gamma^*$ momentum $k+q$ and the light quark
momentum $u-k$. So, the variables $\alpha_k$ and $\beta_k$ satisfy the following
equations:
\ba
\nonumber
\alpha_kM^2-k_T^2=\alpha_k(1+\alpha_k)M_X^2,\\
\beta_k=\frac{M^2-\alpha_kM_X^2}{s},
\ea
where $k_T$ is the transverse momentum of the virtual photon $\gamma^*$,
$M^2$ is the virtuality of the photon $\gamma^*$ (the invariant mass
squared of the muon pair), and $M_X^2$ is the invariant mass squared of the
diffractive final system (including the muon pair and the light quark jet), i.e.
\beq
\label{e1}
M_X^2=(q+u)^2.
\eeq

Because we derive the differential cross section $d\sigma/dt$ at $t=0$,
in the calculations the momentum transfer squared of the diffractive
process $qp\to \gamma^*qp$ is set to be zero, i.e., $u^2=t=0$.
This identity, together with the above equation Eq.~(\ref{e1}), gives
the Sudakov variables for momentum $u$ as,
\beq
\alpha_u=0,~~~\beta_u=\frac{M_X^2}{s},~~~\vec{u}_T^2=0.
\eeq

Similarly, for the momentum $l$, the Sudakov variables are
\ba
\nonumber
\alpha_l&=&-\frac{l_T^2}{s},\\
\nonumber
\beta_l&=&\frac{2(k_T,l_T)-(1-\beta_k)l_T^2}{\alpha_ks},~~{\rm for~Diag.~1,~2},\\
\beta_l&=&-\frac{M_X^2-l_T^2}{s},~~{\rm for~Diag.~3,~4},
\ea
where $(k_T,l_T)$ is the 2-dimensional product of the vector $\vec{k_T}$
and $\vec{l_T}$.
We can see that $\beta_l$ is not the same for these four diagrams,
as it is in the case of the diffractive charm jet and $\jpsi$ production
processes\cite{psi,charm}.

Among these variables, there are two small parameters,
\beq
\beta_u\ll 1,~~~~\frac{l_T^2}{M_X^2}\ll 1,
\eeq
which are the basic expansion parameters in the following
calculations.

With the variables introduced above, the differential 
cross section formula for the partonic process $qp\to \gamma^*q p$ can be
written as
\beq
\label{xs}
\frac{d\hat{\sigma}}{dt}|_{t=0}=\frac{dM_X^2dk_T^2d\alpha_k}{16 s^216\pi^3M_X^2}
        \delta(\alpha_k(1+\alpha_k)+\frac{k_T^2-\alpha_kM^2}{M_X^2})\sum \overline{|{\cal A}|}^2,
\eeq
where ${\cal A}$ is the amplitude of the process $qp\to \gamma^*qp$.
We know that the real part of the amplitude ${\cal A}$ is zero,
and the imaginary part of the amplitude ${\cal A}(gp\to \gamma^* qp)$ for each diagram
of Fig.2 has the following general form,
\beq
\label{ima}
{\rm Im}{\cal A}=C_F\int \frac{d^2l_T}{(l_T^2)^2}F\times\bar u
        (u-k)\Gamma_\mu u(q),
\eeq
where $C_F=2/9$ is the color factor for each diagram.
$\Gamma_\mu$ represents some $\gamma$ matrices including one propagator.
$F$ in the integral is defined as
\beq
F=\frac{3}{2s}g_s^2ee_qf(x',x^{\prime\prime};l_T^2),
\eeq
where
\beq
\label{offd1}
f(x',x^{\prime\prime};l_T^2)=\frac{\partial G(x',x^{\prime\prime};l_T^2)}{\partial {\rm ln} l_T^2},
\eeq
where the function
$G(x',x^{\prime\prime};k_T^2)$ is the so-called
off-diagonal gluon distribution function\cite{offd}.
Here, $x'$ and $x^{\prime\prime}$ are the momentum fractions of the proton
carried by the two gluons.
It is expected that for small $x$, there is no big difference between the off-diagonal and
the usual diagonal gluon densities\cite{off-diag}.
So, in the following calculations, we estimate the production rate by
approximating the off-diagonal gluon density by 
the usual diagonal gluon density, 
$G(x',x^{\prime\prime};Q^2)\approx xg(x,Q^2)$, where $x=x_{I\!\! P}=M_X^2/s$.

In the amplitude Eq.~(\ref{ima}), we see that the leading logarithmic contribution comes from
the terms in $\Gamma_\mu$ which are proportional to $l_T^2$.
So, we can expand $\Gamma_\mu$ in terms of $l_T^2$, and take the leading order
terms ($l_T^2$), and neglect the higher order terms.

Futhermore, we note that in the integral of Eq.~({\ref{ima}) the $l_T^0$ terms in $\Gamma_\mu$
coming from all diagrams must be canceled out by each other.
Otherwise, their net sum (order of $l_T^0$) will lead to a linear singularity
as $l_T^2\to 0$ when we perform the integration over $l_T^2$.
The linear singularity is not proper in QCD calculations.
So, we first examine the behavior of the amplitude at the order of $l_T^0$, i.e., in
the limit of $l_T^2\to 0$.
In this limit, the $\Gamma_\mu$ for each diagram has the following
form,
\ba
\label{e13}
\nonumber
\Gamma_\mu^{(1)}&=&-\Gamma_\mu^{(2)}=\frac{s^2}{M_X^2}\alpha_k\gamma_\mu,\\
\Gamma_\mu^{(3)}&=&-\Gamma_\mu^{(4)}=\frac{s^2}{M_X^2}\gamma_\mu.
\ea
To obtain the above result, we have used the following equations,
\ba
\nonumber
\bar u(u-k)(\not\!u-\not\!k )&=&0,\\
\not\!qv(q)&=&0.
\ea

From Eq.~(\ref{e13}), we can see that the contributions of $\Gamma_\mu$
at the order of $l_T^0$ from the four diagrams are canceled out by each other.
There is no contribution to the amplitude Eq.~(\ref{ima}) at this order.
This is what we expected as mentioned above.

From the above analysis, we see that only after summing up all of the four
diagrams contributions it can give
a gauge invariant amplitude for the partonic process $qp\to q\gamma^* p$.
So, the separation of the diagrams according to Ref.\cite{levin}
is not correct.

At the next order expansion of $\Gamma_\mu$, $l_T^2$, the evaluation is
much more complicated, but nevertheless straightforward.
We first give the expansion result for the propagators.
To the order of $l_T^2$, they are
\ba
\nonumber
g_1&=&\frac{1}{\alpha_kM_X^2},~~~~g_4=\frac{1}{M_X^2},\\
\nonumber
g_2&=&\frac{1}{M_X^2}[1+\frac{1+\alpha_k-\beta_k)l_T^2}{\alpha_kM_X^2}+\frac{4(k_T,l_T)^2}{(\alpha_kM_X^2)^2}-\frac{2(k_T,l_T)}{\alpha_kM_X^2}],\\
g_3&=&\frac{1}{\alpha_k}g_2.
\ea

Apart from the propagator expansion, the $\gamma$ matrices in $\Gamma_\mu$
also contain the $l_T^2$ terms.
They mostly come from the Sudakov variables (expressed in $l_T^2$)
of momentum $l$, i.e., $\alpha_l$, $\beta_l$,
and the slasher $\not\! l_T$.
Furthermore, the $2$-dimensional product
$(k_T,l_T)$ also contributes $l_T^2$ terms. After integrating over the
azimuth angle of $\vec{l_T}$, we get the following result,
\ba
\nonumber
\int d^2l_T(k_T,l_T)^2&=&{\pi\over 2}\int dl_T^2k_T^2 l_T^2,\\
\int d^2l_T(k_T,l_T)\!\not l_T&=&{\pi\over 2}\int dl_T^2\!\not k_T l_T^2.
\ea

To simplify our calculations, the contributions from the four diagrams
are added together, and decomposed into several terms as follows.

The terms (defined as $a$-term) coming from the slasher $\not\!l_T$ in the
$\gamma$ matrices, for all these four diagrams to being summed up together, are
\beq
\Gamma_\mu^{(a)}=-\frac{l_T^2}{M_X^2}\frac{1+\alpha_k}{\alpha_k}
        \not \!p\gamma_\mu\not \!p.
\eeq
The terms ($b$-term) from $\alpha_l=-\frac{l_T^2}{M_X^2}$ are
\beq
\Gamma_\mu^{(b)}=\frac{l_T^2}{M_X^2}[(1+\alpha_k)s\gamma_\mu-
        \frac{1+\alpha_k}{\alpha_k}\not \!p\not \!q\gamma_\mu].
\eeq
The terms ($c$-term) from the propagator expansion are
\beq
\Gamma_\mu^{(d)}=-\frac{l_T^2}{(M_X^2)^2}(1+\alpha_k)s^2(\frac{1+\alpha_k-\beta_k}{\alpha_k}
        +\frac{2k_T^2}{\alpha_k^2M_X^2})\gamma_\mu.
\eeq
The last terms come from those which are proportional to  ``$(k_T,l_T)\not\!l_T$".
They are ($d$-term)
\beq
\Gamma_\mu^{(e)}=\frac{l_T^2}{(M_X^2)^2}\frac{1+\alpha_k}{\alpha_k^2}
        (\alpha_ks\gamma_\mu\not\! k_T\not\!p-s\not\!p\not\!k_T\gamma_\mu).
\eeq

Adding up all of the above $a,~b,~c,~d$ terms, 
we get the amplitude squared for the diffractive
process $qp\to \gamma^*q p$ as, after averaging over the spin and
color degrees of freedoms,
\beq
\sum \overline{|{\cal A}|}^2=\frac{16^2\alpha_s^2\alpha e_q^2\pi^5}{9}\frac{s^2M^2}{(M_X^2)^6}
        (\frac{1+\alpha_k}{\alpha_k})^2[2(1+\alpha_k)M^2M_X^2-\alpha_k(M_X^2)^2-2(M^2)^2]
        (xg(x,Q^2))^2.
\eeq
To obtain the above result, we have only taken the leading order contribution,
and neglected the higher order contribution which is proportional to $\beta_u=M_X^2/s$.
The factorization scale in the gluon density is very important because
we know that the parton distributions at small $x$ change rapidly with
this scale.
In Ref.\cite{levin}, they used different scales for their two parts of
the diagrams for the partonic process. However, as discussed in the above calculations,
all of the four diagrams must be summed together to give a gauge invariant
amplitude for the coherent
diffractive production at hadron colliders. So, the scales of the
four diagrams must be the same, for which we choose it to be $Q^2$.
The separation in \cite{levin} is not gauge invariant.

Finally, we get the differential cross section for the partonic process
$qp\to \gamma^*qp$ in the LLA of QCD,
\ba
\label{xs1}
\nonumber
\frac{d\hat{\sigma}}{dt}|_{t=0}&=&dM_X^2dk_T^2d\alpha_k\frac{\pi^2\alpha_s^2\alpha e_q^2}{9}
        [xg(x,Q^2)]^2(\frac{1+\alpha_k}{\alpha_k})^2\\
        &&\frac{M^2[2((1+\alpha_k)M^2M_X^2-\alpha_k(M_X^2)^2-2(M^2)^2]}{(M_X^2)^6\sqrt{(M_X^2-M^2)^2-4k_T^2M_X^2}}
        [\delta(\alpha_k-\alpha_1)+\delta(\alpha_k-\alpha_2)],
\ea
where $\alpha_{1,2}$ are the solutions of the following equations,
\beq
\alpha(1+\alpha)+\frac{k_T^2-\alpha M^2}{M_X^2}=0.
\eeq

With this differential cross section formula for the partonic process
$qp\to \gamma^*qp$, we can get the differential cross section for the
partonic process of massive muon pair production $qp\to \mp qp$,
\beq
\frac{d\hat{\sigma}(qp\to \mp qp)}{dM^2dt}|_{t=0}=
\frac{\alpha}{3\pi M^2}\frac{d\hat{\sigma}(qp\to \gamma^* qp)}{dt}|_{t=0}.
\eeq

For the $W^\pm$ boson production processes, the differential cross section
formula $d\sigma/dt$ can be easily obtained from Eq.~(\ref{xs1}) by making the
following replacements,
\beq
M\to M_W,~~~\alpha e_q^2\to 2\frac{g_w^2}{4\pi},
\eeq
where $g_w=G_FM_W^2/\sqrt{2}$.

\section{Numerical results}

With the cross section formulas given in last section, we
can calculate the cross section of the diffractive production
at the hadron level.
However, as mentioned above, there exist nonfactorization effects caused by
the spectator interactions in the hard 
diffractive processes in hadron collisions.
Here, we use a suppression factor ${\cal F}_S$ to describe this
nonfactorization effects in the hard diffractive processes at hadron
colliders\cite{soper}.
At the Tevatron,
the value of ${\cal F}_S$ may be as small as ${\cal F}_S\approx 0.1$\cite{soper,tev}.
That is to say, the total cross section of the diffractive processes
at the Tevatron may be reduced down by an order of magnitude due to
the nonfactorization effects.
In the following numerical calculations, we adopt this suppression factor value
to evaluate the diffractive production rates of massive muon pair and $W$ boson
at the Fermilab Tevatron.

In the numerical calculations, we take the input parameters as follows,
\beq
\alpha=1/128,~~~M_W=80.33~GeV,~~~G_F=1.15\times 10^{-5}~GeV^{-2}.
\eeq
The scales for the running coupling constant and the parton distribution
function are set to be the same. For the numerical calculations,
the scale is set to be $Q^2=M^2+k_T^2$ for massive muon pair production
(here we still use $\alpha=1/128$ for this process approximately)
and $Q^2=M_W^2$ for the $W$ boson production.
For the parton distribution functions, we select the GRV NLO set\cite{grv}.

The numerical results for the massive muon pair production at large transverse
momentum in the coherent diffractive processes at the Fermilab Tevatron are
shown in Fig.3 and Fig.4.
In Fig.3, we plot the double differential cross section $d^2\sigma^{\mu\mu}/dM^2dt|_{t=0}$
as a function of the lower bound of the transverse momentum of the muon
pair $k_{T{\rm min}}$, where we set $M^2=5~GeV^2$.
In Fig.4, we plot the double differential cross section as a function of the
lower bound of the momentum fraction $x_1$, $x_{1{\rm min}}$.
This figure shows that the dominant contribution to the cross section comes
from the region of $x_1\sim 10^{-1}$. Comparing this result with that of
the diffractive $\jpsi$ and charm jet production cross sections, we find
that the dominant contribution region of $x_1$ to the cross section
of muon pair production here is some orders of magnitude larger than that of
$\jpsi$ and charm jet production.
This is understandable, because the cross sections for the diffractive
processes at hadron colliders depend on the parton distribution
functions as the following form,
\beq
\label{xsb}
d\sigma\propto f_q(x_1,Q^2)(g(x,Q^2))^2,
\eeq
where $x_1$ is the longitudinal momentum fraction of the proton (or antiproton)
carried by the incident parton (quark or gluon).
In the diffractive $\jpsi$ and charm jet production processes, the incident parton
is the gluon, so the cross section is sensitive to the small-$x$ gluon distribution
function in the proton.
In the diffractive massive muon pair and $W$ boson production processes
calculated here, the incident parton is the quark, so the cross section
is sensitive to the large-$x$ quark distribution in the proton.
However, the dependence of the cross sections
on the second factor $(g(x,Q^2)^2$ is the same for these
two types of processes (the quark induced and the gluon induced processes).
This feature is due to the two-gluon exchange model calculation for the diffractive
processes.

The numerical results on the diffractive $W$ boson production at the Fermilab
Tevatron are plotted in Fig.5 and Fig.6. In these two figures, we only plot
the diffractive $W^+$ production cross section
in the process $p\bar p\to W^+\bar p X$.
$W^-$ production in ($p\bar p\to W^-\bar p X$) process and $W^\pm$ production
in ($p\bar p\to W^\pm p X$) can be obtained similarly.
In Fig.5, we plot the differential cross section $d\sigma^W/dt|_{t=0}$ as
a function of the lower bound of the transverse momentum $k_{T{\rm min}}$.
In Fig.6, we plot the dependence of the cross section on $x_{1{\rm min}}$,
where we set $k_{T{\rm min}}=5~GeV$.
From this figure, we see that the dominant contribution comes from the region
of $x_1>10^{-1}$. This is because the large mass of the $W$ boson requires the
value of $x_1$ larger than that for the low mass muon pair production.

For the possibility about the experimental measurement of the coherent
diffractive production, we think that some other mechanisms may be
important in perturbative QCD beyond the two-gluon exchange model, such
as quark pair exchange. We note that quark pair exchange is
suppressed for the case of valence quarks by the factor
$\beta_u=M^2_X/s$ and corresponds to the secondary Reggeons
($\rho,~\omega,~f, ~\ldots$) exchange.
So, the `valence' and `sea' quark contributions may be different
for the quark pair exchange processes.
Work along this way is in preparation.
Also, the survival factor and the off-diagonal parton distribution
function may cause uncertainties as well.
So, we will leave the discussions about the experimental measurement in
a forthcoming paper.

\section{Conclusions}

In this paper, we have derived the formula for the quark initiated coherent
diffractive processes at hadron colliders in perturbative QCD by using the
two-gluon exchange parametrization of the Pomeron model.
We have shown that production cross sections of massive muon pair and $W$ boson
are related to the off-diagonal gluon distribution and large-$x$ quark distribution
in the proton (antiproton).
To estimate the production rates for these processes, we have approximated the
off-diagonal gluon distribution by the usual gluon distribution function in the proton.

By now, we have studied the coherent diffractive processes at hadron colliders
by using the two-gluon exchange model including the $\jpsi$ production\cite{psi},
the charm jet production\cite{charm}, and the massive muon pair and $W$ boson
productions.
All of these processes have a common feature: there is a large energy scale associated
with these processes to guarantee the application of perturbative QCD,
i.e., $M_\psi$ for $\jpsi$ production, $m_c$ for the charm jet
production, $M^2$ for the massive muon pair production,
and $M_W$ for the $W$ boson production.
And another common feature is the sensitivity to the off-diagonal gluon
distribution function in all these processes (for detailed discussion see
\cite{charm}).
Therefore, these diffractive production processes will open a useful window
for the study of the off-diagonal gluon distribution function in the proton.

\acknowledgments
This work was supported in part by the National Natural Science Foundation
of China, the State Education Commission of China, and the State
Commission of Science and Technology of China.


\newpage
\vskip 10mm
\centerline{\bf \large Figure Captions}
\vskip 1cm
\noindent
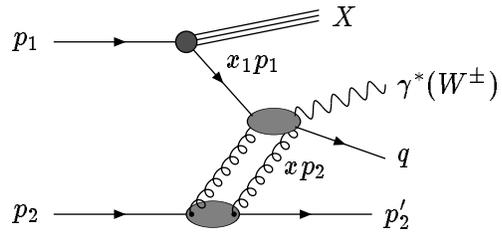
Fig.1. Sketch diagram for the diffractive virtual photon $\gamma^*$ and $W$ boson productions
at hadron colliders in perturbative QCD. 

\noindent
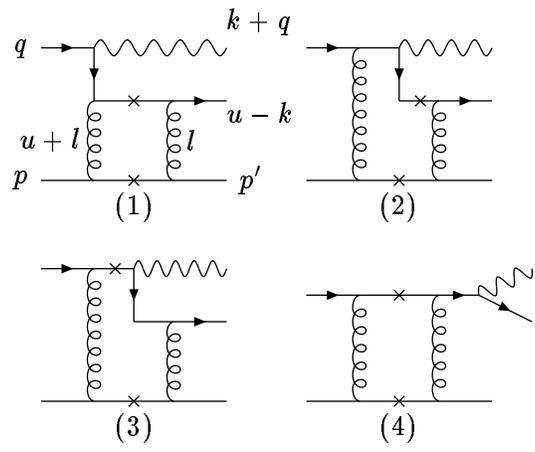
Fig.2. The lowest order perturbative QCD diagrams for partonic process
$qp\to \gamma^* qp$.
The crosses in each diagram represent the discontinuity we calculated in
the evaluation of the imaginary part of the amplitude.

\noindent
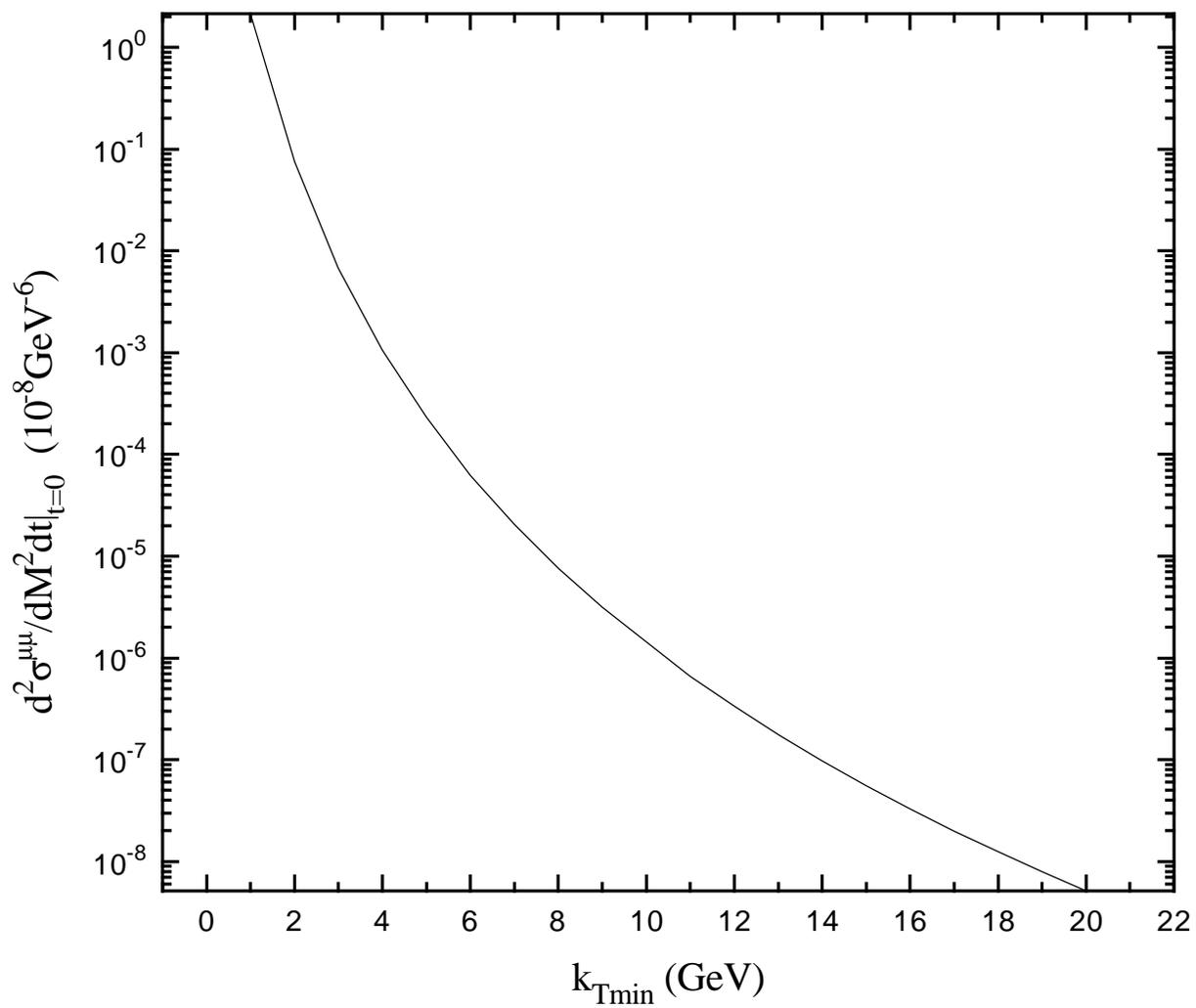
Fig.3. The double differential cross section $d^2\sigma^{\mu\mu}/dM^2dt|_{t=0}$
at $M^2=5~GeV^2$ for the massive muon pair production at the Fermilab Tevatron as a
function of $k_{T{\rm min}}$. $k_{T{\rm min}}$ is the lower bound of the
transverse momentum of the muon pair.

\noindent
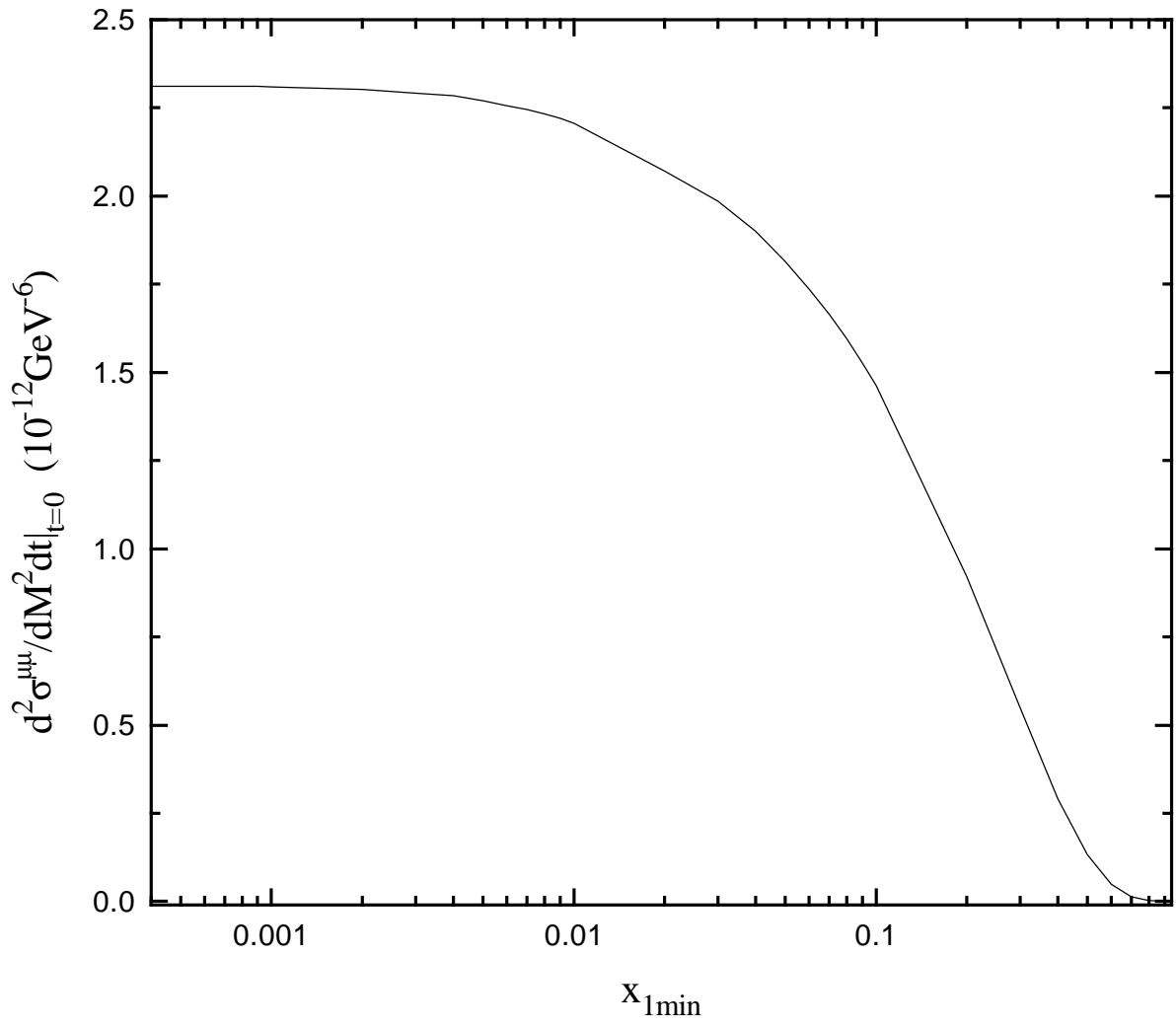
Fig.4. The double differential cross section $d^2\sigma^{\mu\mu}/dM^2dt|_{t=0}$
at $M^2=5~GeV^2$ as a function of $x_{1{\rm min}}$, where we set $k_{T{\rm min}}=5~GeV$.
$x_{1{\rm min}}$ is the lower bound of the momentum fraction of the proton
carried by the incident quark.

\noindent
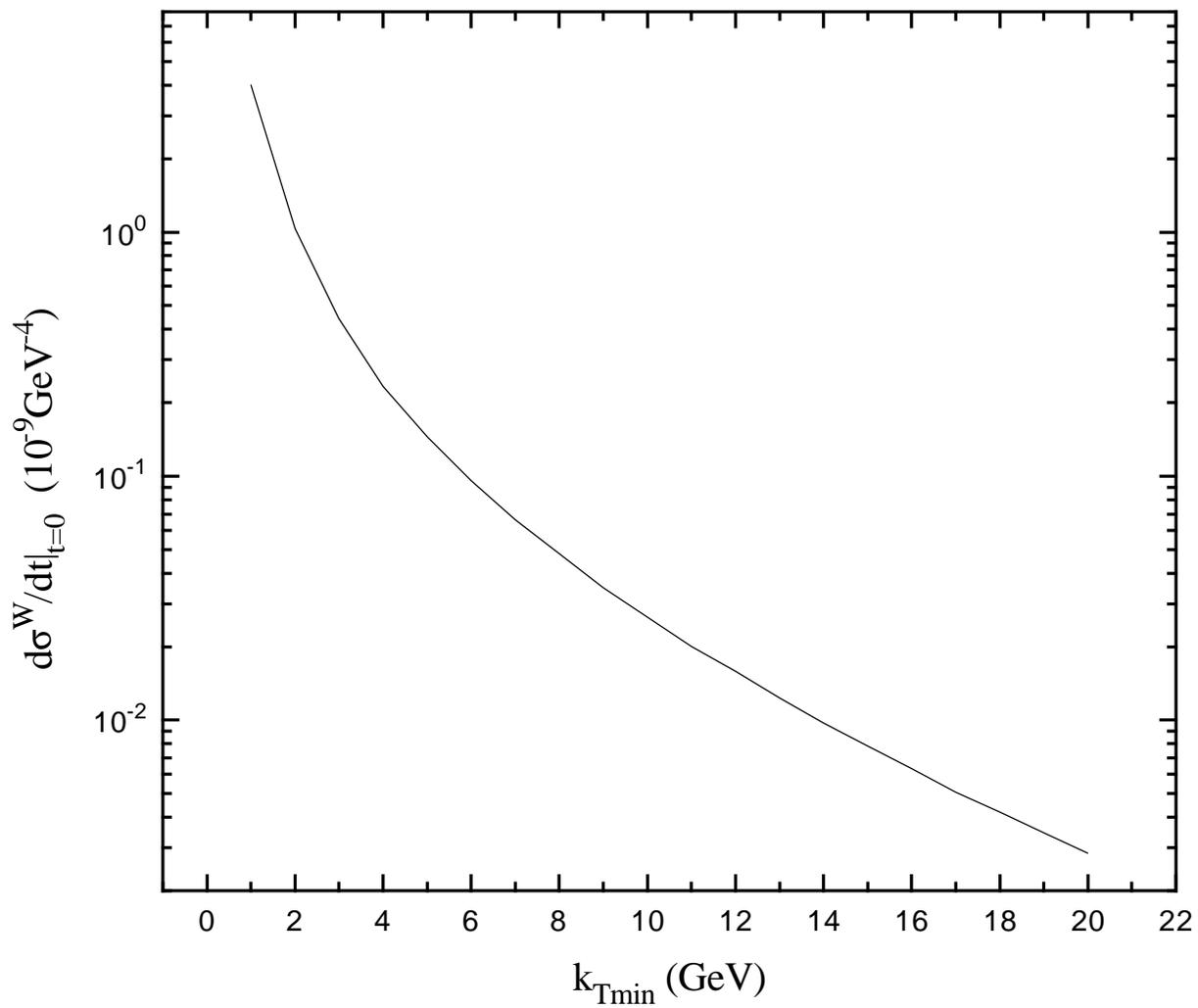
Fig.5. The differential cross section $d\sigma^W/dt|_{t=0}$ for the $W^+$ boson
production in the diffractive process $p\bar p\to W^+\bar p X$ at the Fermilab
Tevatron as a function of $k_{T{\rm min}}$.

\noindent
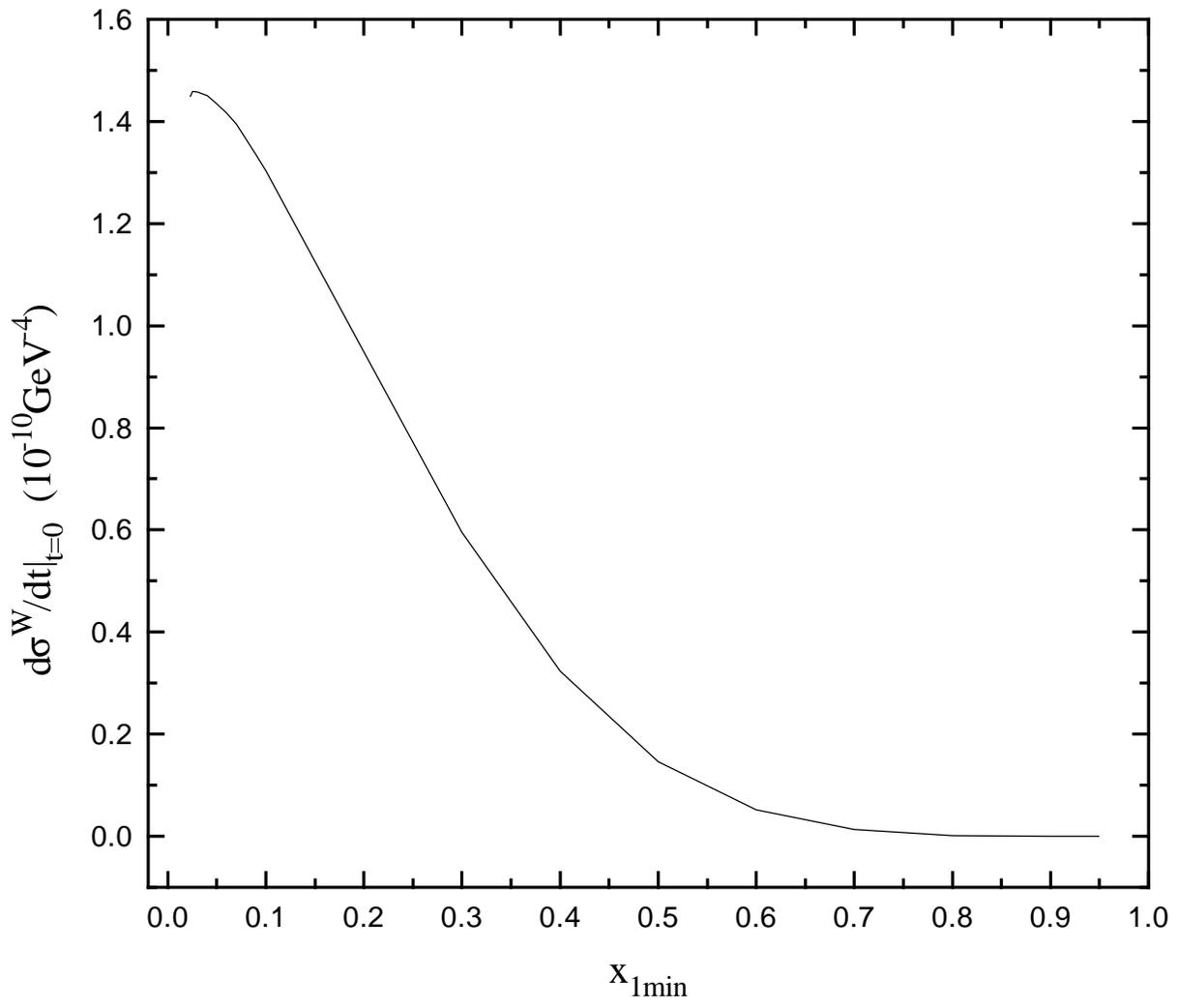
Fig.6. The differential cross section $d\sigma^W/dt|_{t=0}$
as a function of $x_{1\rm min}$, where we set $k_{T{\rm min}}=5~GeV$.

\end{document}